\documentclass[10pt]{article}
\usepackage{amsmath}
\usepackage{amsfonts}
\usepackage{amssymb}

\newcommand{\D}{\,\mathrm{d}}
\newcommand{\E}{\mathrm{e}}
\newcommand{\pd}[2]{\ensuremath{\frac{\partial #1}{\partial #2}}}
\newcommand{\pdl}[2]{\ensuremath{\partial #1 / \partial #2}}
\newcommand{\pdc}[3]{\ensuremath{\left(\frac{\partial #1}{\partial #2}\right)_{#3}}}
\newcommand{\ie}{\textit{i.e.},~}

\author{L.A.\,Melnikovsky\footnote{E-mail: leva@kapitza.ras.ru}}
\title{On the Bose-Einstein Condensation of Rotons}
\date{}
\begin{document}
\maketitle
\begin{center}
\textit{P.L.\,Kapitza Institute for Physical Problems}\\
\textit{Russian Academy of Sciences, 119334 Moscow, Russia}
\end{center}

\begin{abstract}
Bose-Einstein condensation of rotons in helium was
considered long ago. It was shown that the relative velocity of the
normal motion in this state must be equal to the Landau critical
velocity. We argue that the condensation can be attained at a
smaller velocity if the temperature is low enough.

%
%

\end{abstract}

\section{Introduction}
As a Bose gas is cooled down, the Bose-Einstein condensation (BEC)
emerges to avoid a conflict between the statistics and the particle
conservation requirement. Behaviour of quasi-particles, unlike behaviour
of real particles, is not dominated by such conflict: their number is
not an independent variable, it adjusts itself to maximize entropy in
equilibrium. Conventional Bose-Einstein condensation of quasiparticles
is therefore thought to be impossible, an ingenious
mechanism~\cite{jordan} of the condensation for rotons via gap
cancellation by a critical ($v_n-v_s \equiv v =v_L \equiv \Delta_0/P_0$,
where $\Delta_0$ is the roton energy gap and $P_0$ is its momentum)
superfluid counterflow is required\footnote{According to
G.E.\,Volovik~\cite{volovik}, the possibility of the roton BEC was
first investigated (unpublished) by him as early as 40 years ago.}.

Actually, even real particles hardly ever conserve exactly. Consider
atoms of a cold gas in a trap: they can combine into molecules or even
evaporate from the trap altogether. Nor particle conservation is
present in exact relativistic theory. The BEC is still possible if it
forms much faster than the particle population decays. It is therefore
necessary to compare all relevant thermalization time scales. It turns
out that any finite counterflow and low enough temperature $T \ll v P_0$
are the conditions favourable for the roton BEC creation.

Suppose these conditions are satisfied, and the roton number relaxation
is much slower than that of their energy and momentum. The roton
distribution is then characterized by the temperature, velocity, and
finite chemical potential~$\mu$:
\begin{equation}
\label{bose}
N_\mathbf{P}=\left(\exp \frac{\mathcal{E}-\mathbf{Pv}-\mu}{T}-1\right)^{-1},
\end{equation}
where $\mathbf{P}$ is the roton momentum and 
$\mathcal{E}=\Delta_0+(P-P_0)^2/(2\mu_0)$ is its energy. 
Let the $z$ axis run along $\mathbf{v}$ direction.
The roton distribution argument can be expanded in powers of small
deviation from its most probable value
\begin{multline}
\label{eofg}
\mathcal{E}-\mathbf{vP}-\mu \approx
 \Delta_0-\mu-P_0 v -\mu_0 v^2 +
 \frac{f_z^2}{2\mu_0} +
 \frac{\left(f_x^2+f_y^2\right) v}{2P_0}
 \equiv \\
 \Delta +
 \frac{f_z^2}{2\mu_0} +
 \frac{\left(f_x^2+f_y^2\right) v}{2P_0}
=
 \Delta +
 \frac{g^2}{2\mu_0},
\end{multline}
where $\mathbf{f}=\mathbf{P}-\mathbf{v}\left(\mu_0+P_0/v\right)$,
\begin{equation}
\label{redmom}
\begin{aligned}
g_{x,y} &= f_{x,y} \sqrt{\mu_0 v / P_0},\\
g_z &= f_z.
\end{aligned}
\end{equation}
This expansion is applicable if
\begin{equation}
\label{mv2}
T\ll\mu_0 v^2.
\end{equation}


Whether the roton BEC state is a Bogolyubov-like gas or a degenerate Bose
liquid depends on the concentration. Critical concentration (particles
per unit volume) required for Bose-Einstein condensation is
\begin{equation}
\label{CC}
N_c=\zeta\left(\frac{3}{2}\right)
\left(\frac{T}{2\pi}\right)^{3/2} \frac{\mu_0^{1/2} P_0}{v\hbar^3}.
\end{equation}
The system is almost ideal (see \cite{LL9}) if
\begin{equation}
\label{Ca}
N \frac{V_0^3 \mu_0 P_0^2}{2^6 \pi^3 \hbar^6 v^2}
\ll 1,
\end{equation}
where \cite{lk1} $V_0 \sim 10^{-38}\,\mathrm{erg}\,\mathrm{cm}^3$ is the interaction strength.
Combining \eqref{CC} and \eqref{Ca} we get
\[
T
\ll
\frac{2^{5} \pi^3 \hbar^{6} }{\zeta(3/2)^{2/3} V_0^2 \mu_0 P_0^2} v^2
\sim
10^3 \mu_0 v^2.
\]
This inequality is satisfied as a consequence of \eqref{mv2}, implying
that the condensation considered is a transition between gaseous phases.

\section{Relaxation}
\subsection{Roton number decay}
Suppose initial roton distribution is characterized by some positive
chemical potential $\mu$. This means that the roton number is greater
than that in complete equilibrium. The most important process at low
temperature $T \ll v P_0$ for the chemical potential relaxation is the
transformation of two rotons into one roton and one phonon. Little is
known about the transformation probability in such collisions. It seems
reasonable to assume that the transformation cross section (providing
the process is allowed at all by the conservation laws) can be bounded
from above by complete scattering cross section \cite{lk1} known from the
experimental viscosity data.

Momentum conservation for this transformation imposes severe restriction
on the angle $\phi$ between momenta of the incident rotons. Namely, this
restriction is reduced to
$\phi \gtrsim 2 \pi/3$
if the inequality $\Delta \ll P_0 c$ is taken into account. Here $c$ is
the speed of sound.
To simplify all assessments below we take $T\ll \Delta$, \ie assume
Boltzmann statistics for the rotons:
\begin{equation}
\label{boltzmann}
N_\mathbf{P}=\exp \frac{\mathbf{Pv}+\mu-\mathcal{E}}{T}.
\end{equation}
This is certainly incorrect for the BEC state itself but must provide
reasonable relative order of magnitude for different relaxation rates.


As an estimate, not more than about $\exp(-2(1-\cos\pi/3)vP_0/T)$ fraction
of all collisions end up with the transformation.
For the chemical potential relaxation rate this gives (see \cite{lk1})
\begin{equation}
\label{murelax}
\tau^{-1}_{\mu} \lesssim
\frac{4 N |V_0|^2 P_0 \mu_0}{\hbar^4}
\exp \frac{-vP_0}{T} ,
\end{equation}
where $N$ is the total roton concentration. The relaxation rate here is
defined according to
\[
\dot{\mu} + \tau^{-1}_{\mu}\mu=0.
\]


\subsection{Phonon-roton velocity relaxation}
To find the upper boundary for the phonon-roton relaxation time it is
sufficient to consider the two-particle scattering of rotons by phonons.
Conservation laws for this process are (primes denote the finite state)
\begin{equation}
\label{conserv}
\begin{aligned}
\mathbf{p}+\mathbf{P}&=\mathbf{p}'+\mathbf{P}',\\
\varepsilon+\mathcal{E}&=\varepsilon'+\mathcal{E}',
\end{aligned}
\end{equation}
where $\mathbf{p}$ and $\varepsilon=cp$ are the phonon momentum and
energy. Since $P \gg p$, we conclude that $p\approx p'$ and
$\mathbf{P}\parallel\mathbf{P}'$.

Let the roton and phonon subsystems be separately in equilibrium. The
roton and the phonon ``bath'' velocities will be $\mathbf{v}$ and
$\mathbf{v}+\delta \mathbf{v}$ respectively. Velocity relaxation has
two distinct time scales $\tau_{ph, v \parallel}$ and $\tau_{ph, v
\perp}$, they correspond to $\delta\mathbf{v} \parallel \mathbf{v}$ and
to $\delta\mathbf{v} \perp \mathbf{v}$. In linear approximation ($\delta
v \ll v$) the relaxation is described by the equations
\begin{align*}
\dot{j}_{r\parallel} &= \delta v_\parallel \rho_{r \parallel}\tau^{-1}_{ph, v \parallel} ,\\
\dot{j}_{r\perp}     &= \delta v_\perp \rho_{r \perp}\tau^{-1}_{ph, v \perp},
\end{align*}
where $\mathbf{j}$ is the roton momentum density and $\rho_r$ is the
roton contribution to the normal density $\rho_n=\pdl{j}{v}$
\begin{equation}
\label{rho}
\begin{aligned}
\rho_{r \parallel} &= \mu_0 N,\\ 
\rho_{r \perp} &= P_0 N / v.
\end{aligned}
\end{equation}
The relaxation process is governed by the kinetic equation, in the spatially uniform
case it gives
\begin{equation}
\label{jd1}
\dot{\mathbf{j}}_r =
\int (\mathbf{P}-\mathbf{P}')
		(n'+1) n N_\mathbf{P} \,
                \D w \,
                \frac{\D\mathbf{p}}{(2\pi\hbar)^3}
                \frac{\D\mathbf{P}}{(2\pi\hbar)^3},
\end{equation}
where 
\[
n=\left(\exp \frac{\varepsilon-\mathbf{p}(\mathbf{v}+\delta\mathbf{v})}{T}-1\right)^{-1},
\]
the scattering probability is $\D w = c \D\sigma$ and differential
cross-section according to Ref.\cite{lk1} is
\begin{equation*}
\D\sigma=\left(\frac{P_0p^2}{4\pi\hbar^2\rho c}\right)^2
    \left\{
	(\mathbf{n}+\mathbf{n}',\mathbf{m})(\mathbf{n},\mathbf{n}') +
	\frac{P_0}{\mu_0 c}(\mathbf{n},\mathbf{m})^2 (\mathbf{n}',\mathbf{m})^2 +
	A
    \right\}^2 \D\mathbf{n}'.
\end{equation*}
Here $\mathbf{n}$, $\mathbf{n}'$, and $\mathbf{m}$ are the unit vectors
directed along $\mathbf{p}$, $\mathbf{p}'$, and $\mathbf{P}$
respectively. At low temperature most rotons have momentum parallel to
the velocity and the vector $\mathbf{m} \parallel \mathbf{v}$ can be
regarded as a constant. The parameter $A$ is given by
\begin{equation*}
A=\frac{\rho^2}{P_0c}
    \left[
	\pd{^2 \Delta_0}{\rho^2} +
	\frac{1}{\mu_0}\pdc{P_0}{\rho}{}^2
    \right].
\end{equation*}

To transform Eq.\eqref{jd1} we employ the 
usual relation between distribution functions
\[
(n'+1)nN_\mathbf{P} - (n+1)n'N_{\mathbf{P}'} \approx
(n+1)n'N_{\mathbf{P}'} \frac{(\mathbf{P}'-\mathbf{P},\delta \mathbf{v})}{T}
\]
and the principle of detailed balance:
\begin{multline}
\label{jd2}
\dot{\mathbf{j}}_r =
\frac{1}{2T} \int (\mathbf{P}'-\mathbf{P})
(\mathbf{P}'-\mathbf{P},\delta \mathbf{v})
(n+1)n'N_{\mathbf{P}'}
                \D w \,
                \frac{\D\mathbf{p}}{(2\pi\hbar)^3}
                \frac{\D\mathbf{P}}{(2\pi\hbar)^3}=\\
\frac{c}{2T} \int p^2 (\mathbf{n}-\mathbf{n}')
(\mathbf{n}-\mathbf{n}',\delta \mathbf{v})
(n+1)n'N_{\mathbf{P}'}
                \D\sigma \,
                \frac{\D\mathbf{p}}{(2\pi\hbar)^3}
                \frac{\D\mathbf{P}}{(2\pi\hbar)^3}=\\
\frac{\pi^3 P_0^2 T^8}{60 c^{10} \hbar^7 \rho^2}
\int N_\mathbf{P} \frac{\D\mathbf{P}}{(2\pi\hbar)^3}
\int (\mathbf{n}-\mathbf{n}')
(\mathbf{n}-\mathbf{n}',\delta \mathbf{v})
\times \\
\left\{
	\left(\mathbf{n}+\mathbf{n}',\mathbf{m}\right)
	\left(\mathbf{n},\mathbf{n}'\right) +
	\frac{P_0}{\mu_0 c}\left(\mathbf{n},\mathbf{m}\right)^2 
	\left(\mathbf{n}',\mathbf{m}\right)^2 +
	A
\right\}^2 \D\mathbf{n}'
		\D\mathbf{n} 
. 
\end{multline}
For the relaxation rate this gives (lengthy but straightforward
transformations are omitted)
\begin{align}
\label{vparph}
\tau^{-1}_{ph, v \parallel} &=
\frac{8 \pi^5 P_0^2 T^8}{15 c^{10}  \hbar^7 \rho^2 \mu_0}
\left(
\frac{4}{225} + \frac{2 A P_0}{15\mu_0 c} + \frac{P_0^2}{35 \mu_0^2 c^2} + \frac{A^2}{3} 
\right) \sim
500\frac{T^8 P_0^2}{c^{10}  \hbar^7 \rho^2 \mu_0},
\\
\label{vperph}
\tau^{-1}_{ph, v \perp} &=
\frac{8 \pi^5 v P_0 T^8}{15 c^{10} \hbar^7 \rho^2}
\left(
\frac{8}{225} + \frac{2 A P_0}{45\mu_0 c} + \frac{P_0^2}{175 \mu_0^2 c^2} + \frac{A^2}{3} 
\right) \sim
300 \frac{T^8 P_0 v}{c^{10} \hbar^7 \rho^2}
.
\end{align}

\subsection{Phonon-roton temperature relaxation}
Let us now find the temperature equilibration rate between the roton and
the phonon subsystems. The phonon gas temperature is $T+\delta T$ and
the phonon distribution is
\begin{equation*}
n=\left(\exp \frac{\varepsilon-\mathbf{pv}}{T+\delta T}-1\right)^{-1}.
\end{equation*}
This function satisfies equality
\begin{equation*}
(n'+1)nN_\mathbf{P} - (n+1)n'N_{\mathbf{P}'} \approx
(n+1)n'N_{\mathbf{P}'} \frac{\mathcal{E}'-\mathcal{E}}{T^2}\delta T.
\end{equation*}
From the conservation laws \eqref{conserv} it follows that
\begin{equation*}
\mathcal{E}'-\mathcal{E}=
\frac{P-P_0}{\mu_0} \left(\mathbf{m},\mathbf{P}'-\mathbf{P}\right)=
v \left(\mathbf{m},\mathbf{P}'-\mathbf{P}\right)
.
\end{equation*}

The energy inflow to the roton subsystem is
\begin{multline}
\label{ed1}
\dot{E}_r = \int (\mathcal{E}'-\mathcal{E})
		(n'+1) n N_\mathbf{P} \,
                \D w \,
                \frac{\D\mathbf{p}}{(2\pi\hbar)^3}
                \frac{\D\mathbf{P}}{(2\pi\hbar)^3}=\\
\frac{v^2 \delta T}{2T^2}\int (\mathbf{m},\mathbf{P}'-\mathbf{P})^2
		(n+1)n'N_{\mathbf{P}'}
		\D w\,
		\frac{\D\mathbf{p}}{(2\pi\hbar)^3}
		\frac{\D\mathbf{P}}{(2\pi\hbar)^3}
.
\end{multline}
The phonon-roton temperature relaxation rate defined by
\begin{equation*}
\dot{E}_r = C_r \tau_{ph,T}^{-1} \delta T,
\end{equation*}
where $C_r=3N/2$ is the roton contribution to the specific heat per unit
volume, can be immediately extracted using obvious similarity between
\eqref{ed1} and \eqref{jd2}:
\begin{multline}
\label{tph}
\tau_{ph,T}^{-1}=
\frac{v^2 \rho_{r \parallel}}{C_r T}\tau^{-1}_{ph, v \parallel}=
\frac{16 \pi^5 P_0^2 T^7 v^2}{45 c^{10}  \hbar^7 \rho^2}
\left(
\frac{4}{225} + \frac{2 A P_0}{15\mu_0 c} + \frac{P_0^2}{35 \mu_0^2 c^2} + \frac{A^2}{3} 
\right) \sim\\
300\frac{T^7 P_0^2 v^2}{c^{10}  \hbar^7 \rho^2}.
\end{multline}

\subsection{Roton-roton velocity relaxation}
Equilibrium within the roton subsystem is reached via the roton-roton
collisions. Like a two-body problem in classical mechanics
these collisions are efficiently described (if \eqref{eofg} holds)
in the center of inertia frame. Namely, suppose reduced momenta,
defined according to \eqref{redmom}, of the scattering rotons are
$\mathbf{g}_1$ and $\mathbf{g}_2$. After a transformation
\begin{align*}
\mathbf{G} &= \mathbf{g}_1+\mathbf{g}_2,\\
\mathbf{g} &= (\mathbf{g}_1-\mathbf{g}_2) / 2,
\end{align*}
the net energy of two rotons is given by
\begin{equation}
\label{csystem}
\mathcal{E}_1+\mathcal{E}_2-\mathbf{vG}-2\mu=
2\Delta+\frac{G^2}{4\mu_0}+\frac{g^2}{\mu_0}
\end{equation}
and the conservation laws are simplified to
\begin{align*}
\mathbf{G}' &= \mathbf{G}\\
g' &= g.
\end{align*}

Accurate definition of the roton-roton equilibration time constants is
hardly possible (\textit{cf.} discussion on the establishment of
equilibrium of a phonon gas in Ref.\cite{lk2}). As an estimate we employ
the relations similar to those for the phonon-roton relaxation.
\begin{align*}
\rho_{r \parallel} \tau^{-1}_{r, v \parallel} &=
\frac{1}{2T} \int 
\left(\mathbf{P}_1'-\mathbf{P}_1,\mathbf{m}\right)^2
N_{\mathbf{P}_1}N_{\mathbf{P}_2}
    \D w \,
    \frac{\D\mathbf{P}_1}{(2\pi\hbar)^3}
    \frac{\D\mathbf{P}_2}{(2\pi\hbar)^3},\\
\rho_{r \perp} \tau^{-1}_{r, v \perp} &=
\frac{1}{2T} \int 
\left[\mathbf{P}_1'-\mathbf{P}_1,\mathbf{m}\right]^2
N_{\mathbf{P}_1}N_{\mathbf{P}_2}
    \D w \,
    \frac{\D\mathbf{P}_1}{(2\pi\hbar)^3}
    \frac{\D\mathbf{P}_2}{(2\pi\hbar)^3}.
\end{align*}
We follow Ref.\cite{lk1} and adopt the simplest possibility for
the roton-roton scattering probability
\begin{equation*}
\D w =
\frac{8 \pi |V_0|^2}{\hbar}
\delta\left(\mathcal{E}_1+\mathcal{E}_2-{\mathcal{E}_1}'+{\mathcal{E}_2}'\right)
\frac{\D{\mathbf{P}_1}'}{(2\pi\hbar)^3}.
\end{equation*}
For the relaxation time this gives
\begin{multline*}
\rho_{r \parallel} \tau^{-1}_{r, v \parallel} =
\frac{4 \pi |V_0|^2 \E^{-2\Delta/T} }{\hbar T}
    \int \left({g_{1z}}'-g_{1z}\right)^2
    \exp\left(-\frac{g_1^2 + g_2^2}{2 \mu_0 T}\right)\times\\
		\delta\left(\mathcal{E}_1+\mathcal{E}_2-{\mathcal{E}_1}'+{\mathcal{E}_2}'\right)
                \frac{\D{\mathbf{P}_1}'}{(2\pi\hbar)^3}
                \frac{\D\mathbf{P}_1}{(2\pi\hbar)^3}
                \frac{\D\mathbf{P}_2}{(2\pi\hbar)^3}=\\
\frac{4 \pi |V_0|^2 P_0^3 \E^{-2\Delta/T} }{\hbar T \mu_0^3 v^3}
    \int \left({g_{1z}}'-g_{1z}\right)^2
		\exp\left(- \frac{G^2}{4 \mu_0 T} - \frac{g^2}{\mu_0 T}\right)\times\\
                \delta\left(\frac{g^2-g'^2}{\mu_0}\right)
                \frac{\D{\mathbf{g}}'}{(2\pi\hbar)^3}
                \frac{\D\mathbf{g}}{(2\pi\hbar)^3}
                \frac{\D\mathbf{G}}{(2\pi\hbar)^3}=
\frac{ 8 N^2|V_0|^2 P_0 \mu_0^{3/2} T^{1/2}} {3 \pi^{3/2} \hbar^4 v}
\end{multline*}
and
\begin{equation*}
\rho_{r \perp} \tau^{-1}_{r, v \perp} =
\frac{ 8 N^2|V_0|^2 P_0^2 \mu_0^{1/2} T^{1/2}} {3 \pi^{3/2} \hbar^4 v^2}.
\end{equation*}
Here we substituted the roton concentration according to
\begin{equation*}
N=\E^{-\Delta/T}\int
\E^{-g_1^2 / (2T \mu_0)}
\frac{\D\mathbf{P}_1}{(2\pi\hbar)^3}=
\frac{2^{3/2} \pi^{3/2} T^{3/2} \mu_0^{1/2} P_0 }{v (2\pi\hbar)^3}
\E^{-\Delta/T}.
\end{equation*}
Eventually, using \eqref{rho}, we see that the two rates are equal
\begin{equation}
\label{vparr}
\tau^{-1}_{r, v \parallel} =
\tau^{-1}_{r, v \perp} =
\frac{ 8 N |V_0|^2 P_0 \mu_0^{1/2} T^{1/2}} {3 \pi^{3/2} \hbar^4 v}.
\end{equation}

\subsection{Roton-roton temperature relaxation}
Employing the same approach as in \eqref{tph} we express the roton-roton
temperature relaxation rate through the parallel velocity relaxation rate \eqref{vparr}
\begin{equation}
\label{tr}
\tau_{r,T}^{-1}=
\frac{v^2 \rho_{r \parallel}}{C_r T}\tau^{-1}_{r, v \parallel}=
\frac{16 N |V_0|^2 P_0 \mu_0^{3/2} v} {9 \pi^{3/2} \hbar^4 T^{1/2}}.
\end{equation}



\section{Discussion}
As we mentioned above, possibility of the Bose-Einstein condensation of
rotons depends on the relative magnitude of the different relaxation
rates. We begin with the notion that within the roton gas the
temperature relaxation \eqref{tr} is much faster than the velocity relaxation
\eqref{vparr}
\begin{equation*}
\frac{\tau_{r,T}^{-1}}{\tau^{-1}_{r,v}}
=\frac{2 \mu_0 v^2} {3 T} \gg 1.
\end{equation*}
The latter in turn is always faster than the chemical potential
relaxation \eqref{murelax}
\[
\frac{\tau^{-1}_{r,v}}{\tau^{-1}_{\mu}}
=
\frac{2}{3 \pi^{3/2}}
\sqrt{
\frac{T}{\mu_0 v^2}
\exp\frac{2 vP_0}{T}
}
\gg 1.
\]
This completes the proof that a roton subsystem with
an initially narrow momentum distribution around some
undercritical momentum $P$ such that $0<P-P_0<\mu_0\Delta_0/P_0$
must pass through a BEC state.

Whether an initially wide roton distribution can be experimentally
condensed by the ``phonon cooling'' depends not only on the relation
between the rates of the phonon-roton relaxation \eqref{vparph},
\eqref{vperph}, \eqref{tph} and of the roton number decay \eqref{murelax}. The
latter has exponential dependence on the temperature and can in
principle be made arbitrary slow relatively, but at low temperature the
phonon-roton relaxation is very slow itself and may take too long,
therefore demanding a very large experimental cell. To overcome this
difficulty one could try to condense rotons at rest in the laboratory
frame of reference while the superfluid passes through a capillary.
Andreev reflection of rotons \cite{andreev} at low temperature will
protect the distribution width in the roton-wall collisions.

Note that exponentially slow roton number decay (in contrast with the power-law
temperature dependency of other equilibration rates) is a general result
and does not depend on exact roton-roton interaction.

Experimental observation of the roton BEC should be possible by a number of
techniques:
\begin{itemize}
\item the coherent roton quantum state has finite momentum. Upon BEC
formation, bulk helium acquires a spatial inhomogeneity
\cite{pit}, a one-dimensional density wave. The wavelength is the roton
wavelength and the modulation direction is the velocity $\mathbf{v}$
direction. The periodicity should manifest itself as a Bragg peak in the
x-ray scattering experiments. Actually this roton BEC state is a
supersolid as it simultaneously has superfluid and crystalline order.
Note, that one-dimensional crystalline order is not destroyed by
Landau-Peierls fluctuations \cite{peierls,landau} thanks to the true three-dimensional
superfluid order.

\item another option to probe BEC is to explore excitations of the
condensate. The roton second sound is well studied in normal roton systems
\cite{rss} and may be used for the condensate detection.

\item the roton distribution can be measured directly by the quantum evaporation
\cite{evap}. Delta peak in the distribution function would become an
explicit proof of the BEC formation.
\end{itemize}

Finally, let us remark that the stability analysis performed in Ref.\cite{stab} is irrelevant
for the metastable BEC considered in present paper,
because the roton number is fixed.

\section*{Acknowledgements}
I am grateful to A.F.\,Andreev, V.I.\,Marchenko,  and L.P.\,Pitaevskii
for fruitful discussions. The work was supported in parts by RFBR grant
09-02-00567 and RF president program NSh-65248.2010.2.

\end{document}